\def\Roman#1{\uppercase\expandafter{\romannumeral#1}}
\documentclass[a4paper,12pt]{article}
\usepackage{cite}
\usepackage{layout}
\usepackage{amsmath}
\usepackage{amssymb,amsfonts}
\usepackage[mathscr]{eucal}
\usepackage{textcomp}
\usepackage{hyperref}

\title{Equations of  relative equilibria in Yang-Mills theory}

\author{S. N. Storchak}

\author{S. N. Storchak\footnote{E-mail adress: storchak@ihep.ru}\\
\small{ NRC ``Kurchatov Institute'' -- IHEP,}\\
 \small{Protvino, Moscow Region,  142281,  Russia}}

\begin{document}

\maketitle

\begin{abstract}
The equations for finding  relative equilibra in  a pure Yang--Mills gauge theory with  Coulomb gauge fixing are obtained. 
They follow from our previous work on Wong's equations in  gauge theory.
The equations obtained are similar in form to the equations of relative equilibria for finite-dimensional dynamical systems with symmetry. In  both cases, the description of the reduced motion on the orbit spaces of the corresponding principal fiber bundles is performed using dependent coordinates.

\end{abstract}




\section{Introduction}

In our previous article \cite{Wong} we  obtained  Wong's equations in  the pure Yang--Mills gauge theory with the Coulomb gauge fixing. Our derivation was based on the Marsden-Weinstein reduction theory for the dynamical systems with  symmetry. 
 
As is known, in dynamical systems with gauge degrees of freedom, only an implicit description of local dynamics on the orbit space of a gauge group is usually allowed. This is also valid for the most of finite-dimensional dynamical systems with a symmetry if their   motion on the configuration space   can be regarded as occurring on the total space of the principal fiber bundle. 

In order to apply  the  reduction theory to such dynamical systems,  
  we used  a method in which    the description of the local dynamics was given by means of   dependent coordinates. It was done both  for the finite-dimensional systems \cite{Storchak_1,Storchak_11,Storchak_12, Storchak_2} and for the gauge-invariant dynamical systems \cite{Storchak_YM}. 

Our Wong's equations  for a model  finite-dimensional dynamical  system\footnote{This system describes a motion of a particle on a compact Riemanian manifold with a given free isometric smooth action of a compact semisimple Lie group.} (and   for a pure Yang-Mills dynamical system) were  obtained in the form of   geodesic equations in a special coordinate basis. In this basis, known as  the horizontal lift basis, the original Riemannian metric becomes diagonal.
It is important to note that  horizontal Wong's equation are directly related to the classical Yang-Mills equations on the reduced space. To obtain the Yang-Mills equation, it is only necessary to add the potential term to the horizontal Wong equation.
In the present article we will consider the  one of the consequences that follows from  the reduced Yang--Mills equations. It will be shown that  a particular case of these equations can be used to determine   the relative equilibrium in the reduced Yang-Mills system. 

In a classical mechanics, the relative equilibrium of the dynamical system with  symmetry is  such a movement of the system in which the shape of the system does not changed. In other words, the system performs steady motion in a ``group direction'' without violating the shape of the entire  system. Note that the search for the set of points of relative equilibrium is the first step in studying the dynamics of bifurcations in the reduced system. Moreover, after determining the points of relative equilibrium and establishing their stability, we can use an approximate description of the motion of the system in the neighborhood of the equilibrium point.

These and other questions related to the study of various aspects of relative equilibria in finite-dimensional dynamical systems (and also in certain infinite-dimensional systems) have been considered by many authors \cite{Marsd, Marsd_1,1211_5752,1311_7447}. 
In the cited works, additional references to the literature of this direction can be found.

In the first part of our article, we  introduce the notations and definitions, taking them from \cite{Wong}. In addition, we briefly recall some of the facts obtained there. Then it will be shown how to derive the reduced equations and the equations for a relative equilibrium in the finite-dimensional case and for the  pure Yang-Mills theory.
 
\section{Definitions and notations}
The content of this section is taken from our previous article, where one can also find a more detailed discussion of the questions that are omitted here.

The description of the motion of the scalar particle on a compact manifold $\cal P$ with a given 
 free  isometric smooth   action of a semisimple compact Lie group $\cal G$ is performed within the framework of the approach, 
based on the picture of   the fiber bundle, 
in which  $\cal P$ is represented as  the total space
of the principal fiber bundle $\pi : \cal P \to {\cal P}/{\cal G}=\cal M$.

The introduction of the dependent coordinate on this bundle can be carried out as follows.
We  first replace the original coordinates $Q^A$,  given on a local chart of the manifold $\cal P$, by new coordinates 
$(Q^{\ast}{}^A,a^{\alpha})$ ($A=1,\ldots , N_{\cal P},N_{\cal P}=\dim {\cal P} ;{\alpha}=1,\ldots , N_{\cal G},N_{\cal G}=\dim {\cal G}$) that are related to the fiber  bundle $\pi$. 
In order to have a one-to-one correspondence between  the old coordinates $Q^A$ and $(Q^{\ast}{}^A,a^{\alpha})$, the new coordinates $Q^{\ast}{}^A$ must sutisfy the ``gauge'' relations
${\chi}^{\alpha}(Q^{\ast})=0$. 
 If these restrictions imposed on $Q^{\ast}{}^A$  determine the  submanifold $\Sigma $ in the  manifold $\cal P$, then
 we obtain 
   the trivial principal fiber bundle $P({\cal M},\cal G)$ which
is locally isomorphic to the trivial bundle 
$\Sigma\times {\cal G}\to{\Sigma} $. In this case, and also in the case of  locally studying evolution, the coordinates $Q^{\ast}{}^A$ are used to describe the evolution  given on the manifold $\cal M$.

Replacement of  the original coordinate basis $(\frac{\partial}{\partial Q^A})$ by new basis $(\frac{\partial}{\partial Q^{\ast}{}^A},\frac{\partial}{\partial a^{\alpha}})$ leads to a new  representation of  the original metric ${\tilde G}_{\cal A\cal B}(Q^{\ast},a)$ on the manifold $\cal P$: 

\begin{equation}
\left(
\begin{array}{cc}
G_{CD}(P_{\perp})^{C}_{A}
(P_{\perp})^{D}_{B} & G_{CD}(P_{\perp})^
{D}_{A}K^{C}_{\mu}\bar{u}^{\mu}_{\alpha}(a) \\
G_{CD}(P_{\perp})^
{C}_{A}K^{D}_{\nu}\bar{u}^{\nu}_{\beta}(a) & {\gamma }_{\mu \nu },
\bar{u}_\alpha ^\mu (a)\bar{u}_\beta ^\nu (a)
\end{array}
\right),
\label{1}
\end{equation}
 where $K_{\mu}$ are the Killing vector fields for the Riemannian metric 
$G_{AB}(Q)$.
Here $K_{\mu}$,~$G_{CD}$, ${\gamma }_{\mu \nu }$ and $P_{\perp}$ are taken on  the submanifold $\Sigma \equiv\{{\chi}^{\alpha}=0\}$,
${\gamma}_{\mu \nu}=K^{A}_{\mu}G_{AB}K^{B}_{\nu}$ is the metric given on the orbit of the group action, and
${\bar u}^{\alpha}_{\beta}(a)$ (and ${u}^{\alpha}_{\beta}(a)$) are the auxiliary functions for the group $\cal G$.

The projection operator $P_{\perp}(Q^{\ast})$ is defined as 
\[
(P_{\perp})^{A}_{B}=\delta ^{A}_{B}-{\chi}^{\alpha}_{B}
(\chi \chi ^{\top})^{-1}{}^{\beta}_{\alpha}(\chi ^
{\top})^{A}_{\beta},
\]
 where $(\chi ^{\top})^{A}_{\beta}$ is a transposed matrix to the matrix $\chi ^{\nu}_{B}\equiv \frac{\partial \chi ^{\nu}}{\partial Q^B}$, 
$(\chi ^{\top})^{A}_{\mu}=G^{AB}{\gamma}_
{\mu \nu}\chi ^{\nu}_{B}.$ $(P_{\perp})^{A}_{B}$ 
is used to  project the vectors onto the tangent space to the  surface $\Sigma$.

The pseudoinverse matrix ${\tilde G}^{\cal A\cal B}(Q^{\ast},a)$ to the matrix (\ref{1}), i.e.  such a matrix for which 
\begin{eqnarray*}
\displaystyle
{\tilde G}^{\cal A\cal B}{\tilde G}_{\cal B\cal C}=\left(
\begin{array}{cc}
(P_{\perp})^A_C & 0\\
0 & {\delta}^{\alpha}_{\beta}
\end{array}
\right),
\end{eqnarray*}  
is given by 
\begin{equation}
\displaystyle
\left(
\begin{array}{cc}
G^{EF}N^{C}_{E}
N^{D}_{F} & G^{SD}N^C_S{\chi}^{\mu}_D
(\Phi ^{-1})^{\nu}_{\mu}{\bar v}^{\sigma}_{\nu} \\
G^{CB}{\chi}^{\gamma}_C (\Phi ^{-1})^{\beta}_{\gamma}N^D_B
{\bar v}^{\alpha}_{\beta} & G^{CB}
{\chi}^{\gamma}_C (\Phi ^{-1})^{\beta}_{\gamma}
{\chi}^{\mu}_B (\Phi ^{-1})^{\nu}_{\mu}
{\bar v}^{\alpha}_{\beta}{\bar v}^{\sigma}_{\nu}
\end{array}
\right).
\label{2}
\end{equation}
The matrix  $(\Phi ^{-1}){}^{\beta}_{\mu}$  is inverse
to the Faddeev -- Popov matrix $\Phi $,
\[
(\Phi ){}^{\beta}_{\mu}(Q)=K^{A}_{\mu}(Q)
\frac{\partial {\chi}^{\beta}(Q)}{\partial Q^{A}}.
\]
The matrix  ${\bar v}^{\alpha}_{\beta}(a)$ is
an inverse matrix to
 matrix ${\bar u}^{\alpha}_{\beta}(a)$.

The projection operator given by formula
\[
N^{A}_{C}\equiv{\delta}^{A}_{C}-K^{A}_{\alpha }
(\Phi ^{-1}){}^{\alpha}_{\mu}{\chi}^{\mu}_{C}
\]
 has the following properties:
\[
N^{A}_{B}N^{B}_{C}=N^{A}_{C},\,\,\,\,\,N^A_BK^B_{\mu}=0,\,\,\,\,\,
(P_{\perp})^{\tilde A}_{B}N^{C}_{\tilde A}=
(P_{\perp})^{C}_{B},\,\,\,\,\,\,\,N^{\tilde A}_
{B}(P_{\perp})^{C}_{\tilde A}=N^{C}_{B}.
\]

Now, in the next step, one needs to replace the resulting coordinate  basis $(\frac{\partial}{\partial Q^{\ast}{}^A},\frac{\partial}{\partial a^{\alpha}})$ with another basis -- the  horizontal lift basis $(H_A,L_{\alpha})$ \cite{Storchak_3}.  This  
nonholonomic basis consists of the horizontal vector fields $H_A$ and the left-invariant vector fields $L_{\alpha}=v^{\mu}_{\alpha}(a)\frac{\partial}{\partial a^{\mu}}$, satisfying
 the commutation relations
 \[
[L_{\alpha},L_{\beta}]=c^{\gamma}_{\alpha \beta} L_{\gamma},
\]
where the $c^{\gamma}_{\alpha \beta}$ are the structure constants of the group $\cal G$.

The horizontal vector fields $H_A$ are given  as follows
\[
 H_A=N^E_A(Q^{\ast}) \left(\frac{\partial}{\partial Q^{\ast}{}^E}-{\tilde {\mathscr A}}^{\alpha}_E\,L_{\alpha}\right),
\]
where ${\tilde{\mathscr A} }^{\alpha}_E(Q^{\ast},a)={\bar{\rho}}^{\alpha}_{\mu}(a)\,{\mathscr A}^{\mu}_E(Q^{\ast})$. 
The matrix ${\bar{\rho}}^{\alpha}_{\mu}$ is inverse to the matrix ${\rho}_{\alpha}^{\beta}$ of the adjoint representation of the group $\cal G$, and  ${\mathscr A}^{\nu}_P={\gamma}^{\nu\mu}K^R_{\mu}\,G_{RP}$ is the mechanical connection  defined in our principal fiber bundle.

The curvature $\tilde{\mathcal F}^{\alpha}_{EP}$ of the connection ${\tilde{\mathscr A}}$ is given by
\[
\tilde{\mathcal F}^{\alpha}_{EP}=\displaystyle\frac{\partial}{\partial Q^{\ast}{}^E}\,\tilde{\mathscr A}^{\alpha}_P- 
\frac{\partial}{\partial {Q^{\ast}}^P}\,\tilde{\mathscr A}^{\alpha}_E
+c^{\alpha}_{\nu\sigma}\, \tilde{\mathscr A}^{\nu}_E\,
\tilde{\mathscr A}^{\sigma}_P,
\]
($\tilde{\mathcal F}^{\alpha}_{EP}({Q^{\ast}},a)={\bar{\rho}}^{\alpha}_{\mu}(a)\,{\mathcal F}^{\mu}_{EP}(Q^{\ast})\,$).  

We notice that
\[
L_{\alpha}\, {\tilde {\mathscr A}}^{\lambda}_E=-c^{\lambda}_{\alpha \mu}
\,{\tilde {\mathscr A}}^{\mu}_E.
\]
This follows from the equation for ${\rho}$:  $L_{\alpha}\,{\rho}^{\gamma}_{\beta}=c^{\mu}_{\alpha \beta}\,{\rho}^{\gamma}_{\mu}$.

It can be shown that in a new basis in which $[H_A,L_{\alpha}]=0$, the   metric  (\ref{1}) has the following representation: 
\begin{equation}
\displaystyle
{\check G}_{\cal A\cal B}=
\left(
\begin{array}{cc}
G^{\rm H}_{AB} & 0 \\
0 & \tilde{\gamma }_{\alpha \beta }
\end{array}
\right),
\label{metric}
\end{equation}
\[
 {\tilde G}(H_A,H_B)\equiv G^{\rm H}_{AB}(Q^{\ast}), \;\;\;\; 
{\tilde{G}}(L_{\alpha},L_{\beta})\equiv\tilde{\gamma }_{\alpha \beta }(Q^{\ast},a)={\gamma}_{{\alpha}'{\beta}'}(Q^{\ast})\, {\rho}^{{\alpha}'}_{\alpha}(a)\,
{\rho}^{{\beta}'}_{\beta}(a).
\]
The ``horizontal metric'' $G^{\rm H}$ is defined with the help of  
the projection operator ${\Pi}^{ A}_B={\delta}^A_B-K^A_{\mu}{\gamma}^{\mu \nu}K^D_{\nu}G_{DB}$ as follows:  
$G^{\rm H}_{DC}={\Pi}^{\tilde D}_D\,{\Pi}^{\tilde C}_C\,G_{{\tilde D}{\tilde C}}$.
The operator  ${\Pi}^{ A}_B$ has the following properties: ${\Pi}^{ A}_L N^L_C={\Pi}^{ A}_C$ and ${\Pi}^L_BN^A_L=N^A_B$.

The pseudoinverse matrix ${\check G}^{{\mathcal A}{\mathcal B}}$ to the matrix (\ref{metric}) is defined by
the following orthogonality condition:   
\begin{eqnarray*}
\displaystyle
{\check G}^{\mathcal A\mathcal B}{\check G}_{\mathcal B\mathcal C}=\left(
\begin{array}{cc}
N^A_C & 0\\
0 & {\delta}^{\alpha}_{\beta}
\end{array}
\right)
\end{eqnarray*}  
and can be represented as
\begin{eqnarray*}
\displaystyle
{\check G}^{\cal A\cal B}=
\left(
\begin{array}{cc}
G^{EF}N^A_EN^B_F & 0 \\
0 & \tilde{\gamma }^{\alpha \beta }
\end{array}
\right).
\end{eqnarray*}

In new coordinates, the quadratic form of the metric ${\check G}_{\mathcal A\mathcal B}$ is
\[
 ds^2=G^{\rm H}_{AB}\, {\omega}^A{\omega}^B+\tilde{\gamma }^{\alpha \beta }\,{\omega}^{\alpha}{\omega}^{\beta},
\]
where the dual basis to $(H_A,L_{\alpha})$ is given by 
\begin{eqnarray*}
&&{\omega}^A=({\rm P}_{\bot})^A_S\, dQ^{\ast}{}^S,\nonumber\\
&&{\omega}^{\alpha}=u^{\alpha}_{\mu}da^{\mu}+{\tilde{\mathscr A}}^{\alpha}_E \,({\rm P}_{\bot})^E_S\, dQ^{\ast}{}^S,
 \end{eqnarray*}
where $({\rm P}_{\bot})^A_S(Q^{\ast})\, dQ^{\ast}{}^S= dQ^{\ast}{}^A$     and for which we have ${\omega}^{A}(H_B)=N^A_B$,   ${\omega}^{\alpha}(H_A)=0$, and ${\omega}^{\alpha}(L_{\beta})={\delta}^{\alpha}_{\beta}$.
With a new metric ${\check G}_{\mathcal B\mathcal C}$, it is possible to calculate the  Christoffel symbols \cite{Storchak_3} and then to obtain   the geodesic equation \cite{Wong}. 

\section{The equations of motion and relative equilibrium}
For a simple mechanical system with the Lagrangian $L=K-V$ where $K$ is a kinetic energy associated to the obtained Riemannian metric  $ds^2$ and where $V$ is an invariant potential, one can derive, by using the standard methods, the equations of motion.
The horizontal equation (the equation for the horizontal part of motion) is as follows:
\begin{eqnarray} 
&&\frac{d \,{\dot {Q}^{\ast}{}^A}}{\!\!\!dt}
+{}^{\rm H}{\Gamma}^A_{BC}
{\dot {Q}^{\ast}{}^B}{\dot {Q}^{\ast}{}^C}
+G^{AS}N^F_S{\mathcal F}^{\nu}_{EF}{\dot{Q}^{\ast}{}^E}p_{\nu}
+\frac12G^{AS}\,N^E_S\,({ \mathscr D}_E{ \gamma}^{\kappa\sigma})p_{\sigma}p_{\kappa}\nonumber\\
&&\;\;\;\;\;\;\;\;\;\;\;\;\;\;\;\;+G^{AD}{\partial}_D V=0,
\label{4}
\end{eqnarray}
where the variable $p_{\nu}={\gamma}_{\nu \kappa}{\rho}^{\kappa}_{\alpha}z^{\alpha}$  is related to the vertical part of  motion,\footnote{$\,\,\, z^{\alpha}$ is the vertical component of the tangent vector to $\mathcal P.$}
the Christoffel symbols ${}^{\rm H}{\Gamma}^B_{CD}$ are defined by the equality
\begin{eqnarray*}
G^{\rm H}_{AB}\,
{}^{\rm H}{\Gamma}^B_{CD}
=\frac12\left(G^{\rm H}_{AC,D}+
G^{\rm H}_{AD,C}-G^{\rm H}_{CD,A}
\right),
\end{eqnarray*}
 $G^{\rm H}_{AC,D}\equiv\frac{ 
{{\partial G^{\rm H}_{AC}(Q)}}}
{{\partial Q^D}}
|_{Q=Q^{*}}$,
and
the covariant derivative is
 given by
\[
{ {\mathscr D}}_E{ \gamma}_{\alpha\beta}=
\Bigl(\frac{\partial}{\partial Q^{\ast}{}^E}{\gamma}_{\alpha\beta}-c^{\sigma}_{\mu\alpha}{{\mathscr A}}^{\mu}_E{ \gamma}_{\sigma\beta}-c^{\sigma}_{\mu\beta}{{\mathscr A}}^{\mu}_E{ \gamma}_{\sigma\alpha}\,\Bigr).
\]
Notice that the coefficients of the  equation (\ref{4}) depend on ${Q}^{\ast}(t)$. Also note that we  omitted the common multiplier -- the projector $N^B_A$, which stands before the equation. This  operator performs the projection on the plane which is orthogonal to the orbit. 
Moreover, in  deriving the equation, we used the equality $N^E_S\,{\partial}_E V={\partial}_S V$, because $K^A_{\alpha}\partial_A V=0$.  This follows from 
the invariance of the potential $V$  under the action of the group: $V(F(Q,a))=V(Q)$. Therefore, we also have  $V(F({Q}^{\ast},a))=V({Q}^{\ast})$.

The horizontal equation  can be used to describe the motion   on the reduced space. 
The motion of our system on the orbit space of the principal fiber bundle can be investigated by solving this equation for the case $p_{\mu}=0$.
In the theory of reduction, such a case is known as a reduction onto the zero momentum level.
Another special case, when $p_{\mu}=\rm{const}$ in the equation (\ref{4}), is related to the reduction onto the ``non-zero momentum level''. 

It should also be noted that the equation (\ref{4}) is analogous to the corresponding equation obtained in 
 \cite{Littlejohn} for the $n$-body problem in mechanics.

 The vertical  equation of motion (the equation for the internal momentum or  the internal charge) can be written in the following form:  
\begin{equation}
 \frac{d {p}_{\sigma}}{dt}-c^{\kappa}_{\mu \sigma}\,{\mathscr A}^{\mu}_E\,p_{\kappa}\,\dot {Q}^{\ast}{}^E
-c^{\mu}_{\sigma \nu}\,{\gamma}^{\nu \kappa}\,p_{\mu}\,p_{\kappa}=0. 
\label{5}
\end{equation}
This form of the equation is obtained using the following definition for $z^{\alpha}$:
\[
 z^{\alpha}=u^{\alpha}_{\mu}\frac{d a^{\mu}}{dt}+\tilde{\mathscr A}^{\alpha}_E\dot {Q}^{\ast}{}^E.
\]
It is this choice that leads to an intrinsic and global splitting of the original motion on $\mathcal P$ into horizontal and vertical parts\cite{Marsden-Ratiu}.

The relative equilibrium of a dynamical system with symmetry is defined as 
motion for which $\dot {Q}^{\ast}{}^A=0$. (But, in general, $p_{\mu}$ may  be different from zero.)
With this motion, the system may be viewed as a rigid body. 
Setting $\dot {Q}^{\ast}{}^A=0$ in the equation (\ref {4}), we can conclude from it that $p_{\mu}$ is a constant.
Thus, under this substitution, from equations (\ref{4}) and (\ref{5}) we get a system of two equations for the determination of the  relative equilibria:
\begin{eqnarray}
&&\frac12G^{AS}\,N^E_S\,({ \mathscr D}_E{ \gamma}^{\kappa\sigma})p_{\sigma}p_{\kappa}=G^{AE}\,{\partial}_E V,\nonumber\\
&&\;\;\;c^{\mu}_{\sigma \nu}\,{\gamma}^{\nu \kappa}\,p_{\mu}\,p_{\kappa}=0. 
\label{equilibr}
\end{eqnarray}

The second equation of this system is easy solved if we assume that $p_{\mu}$ is an eigenfunction ${\rm e}_{\kappa}$ defined by the matrix equation 
$$ (k_{\varphi \nu}{\gamma}^{\nu \kappa}){\rm e}_{\kappa}=\lambda \,{\rm e}_{\varphi},$$
with $k_{\alpha\beta}=c^{\nu}_{\mu \alpha} c^{\mu}_{\nu \beta}$.

Since ${\gamma}^{\nu \kappa}{\rm e}_{\kappa}=\lambda \,k^{\nu \epsilon}\,{\rm e}_{\epsilon}$, in the second equation of the system we  have $\lambda \,c^{\mu}_{\sigma \nu}k^{\nu \epsilon}\,{\rm e}_{\epsilon}{\rm e}_{\mu}$.  Using the identity $c^{\mu}_{\sigma \nu}k^{\nu \epsilon}=-c^{\epsilon}_{\sigma \nu}k^{\nu \mu}$, we get $c^{\mu}_{\sigma \nu}k^{\nu \epsilon}\,{\rm e}_{\epsilon} {\rm e}_{\mu}=-c^{\epsilon}_{\sigma \nu}k^{\nu \mu}{\rm e}_{\epsilon} {\rm e}_{\mu}$. But this means that 
$c^{\mu}_{\sigma \nu}k^{\nu \epsilon}\,{\rm e}_{\epsilon} {\rm e}_{\mu}=0$,
and therefore the eigenfunction ${\rm e}_{\kappa}$ gives a solution to the second equation of the system.

Having found the solution, we substitute it into the first equation  ({\ref{equilibr}). The resulting equation determines the value of the variable ${Q}^{\ast}{}^A$ at the relative equilibrium of the dynamical system.

\section{The relative equilibrium in Yang--Mills\\theory}
 
In Yang-Mills theory, the initial evolution of the dynamical system occurs  on the functional space of  gauge fields.   
Gauge transformations form a group  acting on this space. 
We assume that this group is the gauge group of time-independent transformations (the gauge condition $A_0=0$ has already been imposed):
\[
{\tilde A}^{\alpha}_i({\mathbf x})={\rho}^{\alpha}_{\beta}(g^{-1}({\mathbf x}))
{ A}^{\beta}_i({\mathbf x})+u^{\alpha}_{\mu}(g({\mathbf x}))
\frac{\partial g^{\mu}({\mathbf x})}
{\partial {\mathbf x}^i}\,,
\]
where ${\rho}^{\alpha}_{\beta}(g)=\bar{u}^{\alpha}_{\nu}(g)\,
v^{\nu}_{\beta}(g)$ is  the matrix of the adjoint
 representation of the group $G$.

To  fix the gauge symmetry, we  use  the Coulomb condition
${\partial}^k A^{\nu}_k({\mathbf x})=0$, 
$\nu=1,\dots,N_{G}$, (or ${\chi}^{\nu}(A)=0$). 
This defines the  gauge surface $\Sigma \equiv\{{\chi}^{\nu}(A)=0\}$.

That is,  the gauge fields $A^{\alpha}_i({\mathbf x})$ now play the role of the original coordinates ``$Q^A$'' 
of a point $p\in \cal P$. 
   
 The Hamiltonian of the pure Yang-Mills theory, which is used in  the Schr\"odinger functional approach is given by
\[
H=\frac 12\mu ^2\kappa \,\triangle
_{\cal P}[A_a]+\frac
1{\mu ^2\kappa }\,
V[A_a],
\]
where 
\[
\triangle _{\cal P}[A]=
\int d^3x \, k^{\alpha \beta}\delta_{ij}
\frac{{\delta}^2}{\delta A^{\alpha}_i({\mathbf x})\;
\delta A^{\beta}_j({\mathbf x})}\,,
\] 
\[
V[A]=\int d^3x \,\frac{1}{2}\,
k_{\alpha \beta}\,
F^{\alpha}_{ij}({\mathbf x})\,F^{\beta \;ij}({\mathbf x})\,.
\]
$k_{\alpha \beta}=c^{\tau}_{\mu \alpha}c^{\mu}_{\tau \beta}$ is the Cartan--Killing metric on the group G, ${\mu}^2=\hbar g_0^2$, and $\kappa$ is a real positive parameter.
Since the quadratic part of the Hamiltonian is as follows
\[
 G^{({\alpha}, i,x)\;({\beta},j,x')}
\frac{{\delta}^2}{\delta A^{(\alpha ,i,x)}\;
\delta A^{(\beta ,j,x')}},\]
 we have a flat metric 
$G^{({\alpha}, i,x)\;({\beta},j,x')}={k}^{\alpha\,\beta}\,
{\delta}^{i\,j}\,{\delta}^3({\mathbf x}-{\mathbf x}').$
The quadratic form of this metric is written as
\[
ds^2=G_{(\alpha , i, x)(\beta,j,y)}\delta A^{(\alpha ,i,x)}
\delta A^{(\beta ,j,y)}\,,
\]
with
\[
G_{(\alpha , i, x)(\beta,j,y)}=G\biggl(\frac{\delta}
{\delta A^{\alpha}_i({\mathbf x})}\,,\,\frac{\delta}
{\delta A^{\beta}_j({\mathbf y})}\biggr)
=k_{\alpha \beta}\delta^{ij}{\delta}^3({\mathbf x}-{\mathbf y})\,.
\]

We must replace the original coordinates 
$A^{\alpha}_i({\mathbf x})$ by new   coordinates \\$(A^{\ast}{}^{\alpha}_i({\mathbf x}),g^{\mu}({\mathbf x}))$. 
 And  $A^{\ast}({\mathbf x})$ must satisfy the  gauge condition: $\chi ^{\alpha}(A^{\ast}({\mathbf x}))=0$.

In order to obtain a functional equation for the motion of Yang-Mills fields, we could follow the same way as in the finite-dimensional case. But the same can be done with the help of our finite-dimensional dynamical system, if we replace the terms in the corresponding equations by their functional analogs.
Instead of finite-dimensional variables, we introduce their functional counterparts  taken from the theory of the gauge fields.
 This means that now we regard the indices of the variables in the finite-dimensional equations as a compact notation of the corresponding extended indices:
\[
A\to(\alpha, i,x);\hspace{2mm}  \mu\to(\mu,u);\hspace{1mm}\ldots\hspace{1mm} {\rm etc}\;.
\]
Then for the time derivative of the basic variable ${Q}^{\ast}{}^B(t)$ we  have the following correspondence:
\[
{\dot {Q}^{\ast}{}^B}(t)\to \frac{d}{dt}{A}{}^{\ast (\sigma,j,y)}(t)\equiv \frac{d}{dt}{A}{}^{\ast \sigma j}({\mathbf y},t)\equiv
{\dot {A}{}^{\ast \sigma j}}({\mathbf y},t).
\]
    Similar substitutions  must be made  in all  variables of  finite-dimensional equations. 
The functional counterparts of the finite-dimensional expressions can be defined in  the standard way. 

Thus,  the Killing vectors are given by 
$K_{(\alpha,y)}=K^{(\mu, i, x)}_{\;\;\;\;\;\;(\alpha,y)}\frac{\delta}
{\delta A^{(\mu ,i,x)}}\,$ with
\[
K^{(\mu, i, x)}_{\;\;\;\;\;\;(\alpha,y)}(A)=
\left[\left({\delta}^{\;\mu}_{\alpha}{\partial}^i({\mathbf x})
+c^{\mu}_{\tilde \nu \alpha}A^{\tilde \nu i}({\mathbf x})
\right){\delta}^3 ({\mathbf x}-{\mathbf y})\right]
\equiv \left[{\mathcal D}^{\mu i}_{\;\;\alpha}(A({\mathbf x}))
\,{\delta}^3({\mathbf x}-{\mathbf y})\right]
\]
(here ${\partial}^i({\mathbf x})$ is a partial derivative 
with respect to $x^i$).
The Killing vectors 
are used for definition of the 
 orbit metric 
\[
\gamma _{(\mu,x)( \nu,y)}=K^{(\alpha,i,z)}_{\;\;\;\;\;\;(\mu,x)}
G_{(\alpha,i,z)(\beta,j,u)}K^{(\beta,j,u)}_{\;\;\;\;\;\;( \nu,y)}\,.
\]

Thus, it is given by
\[
\gamma _{(\mu,x)( \nu,y)}=k_{\varphi \alpha}{\delta}^{kl}
\left[\bigl(-\delta
^{\varphi}_{\,\mu}\;{\partial}_k({\mathbf x})+c^{\varphi}_{\sigma
\mu}A^{\ast}{}^{\sigma}_k({\mathbf x})\bigr)\bigl(\delta
^{\alpha}_{\,\nu}{\partial}_l({\mathbf x})+c^{\alpha}_{\kappa
\nu}A^{\ast}{}^{\kappa}_l({\mathbf x})\bigr){\delta}^3({\mathbf x}-{\mathbf
y})\right].
\]

An "inverse matrix" 
$\gamma ^{( \alpha,y)( \mu,z)}$ to the  matrix $\gamma _{(\mu,x)( \nu,y)}$ 
 can be defined by the following equation:
\[
\gamma _{(\mu,x)( \nu,y)}\;
\gamma ^{( \nu,y)( \sigma,z)}={\delta}^{( \sigma,z)}_
{\;(\mu,x)}\equiv{\delta}^{\sigma}_
{\,\mu}\,{\delta}^3({\mathbf z}-{\mathbf x})\,.
\]
That is,
\[
k_{\varphi \alpha}\,{\delta}^{kl}\,
{\tilde{\cal D}}^{\varphi }_{\mu \,k}(A^{\ast}(\mathbf x))\,
{{\cal D}}^{\alpha }_{\nu \,l}(A^{\ast}(\mathbf x))
\,\gamma ^{( \nu,x)( \sigma,z)}={\delta}^{\sigma}_
{\,\mu}\,{\delta}^3(\mathbf z-\mathbf x)\,.
\]
It follows that $\gamma ^{( \nu,x)( \sigma,z)}$ is the Green function of the operator $({\tilde{\cal D}}\,{{\cal D}})_{\mu \nu}$.
For this, of course, it is also necessary to impose the corresponding boundary conditions.

The mechanical connection ${\mathscr A}^{\alpha}_{\,B}$, known 
in the Yang-Mills fields as  
the ``Coulomb connection'', is given by 
\[
{\mathscr A}^{(\alpha,x)}_{\;\;\;(\beta,j,y)}=
\left[{\cal D}^{{\varphi}}_{\;\mu j}(A^{\ast}({\mathbf y}))
{\gamma}^{(\alpha,x)\,(\mu , y)}\right]\,k_{\varphi \beta}\,.
\]

Since the  Riemannian metric on the original manifold  of the gauge fields is flat, we should  rewrite
 the equation (\ref{4}),  by taking this fact into account. That is,  we should consider the particular case of eq.(\ref{4}), when $G_{AB}={\delta}_{AB}$.
As a result we come to the following horizontal equation:
\begin{eqnarray} 
&&\frac{d \,{\dot {Q}^{\ast}{}^A}}{\!\!\!dt}
+{}^{\rm H}{\Gamma}^A_{BC}
{\dot {Q}^{\ast}{}^B}{\dot {Q}^{\ast}{}^C}
+G^{AS}N^F_S{\mathcal F}^{\nu}_{EF}{\dot{Q}^{\ast}{}^E}p_{\nu}
+\frac12G^{AE}\,({ \mathscr D}_E{ \gamma}^{\kappa\sigma})p_{\sigma}p_{\kappa}\nonumber\\
&&\;\;\;\;\;\;\;\;\;\;\;\;\;\;\;\;+G^{AD}{\partial}_D V=0,
\label{6}
\end{eqnarray}
Using functional representations for the terms of this equation and  performing the necessary generalized summation over repeated indices, we obtain the following horizontal equation of motion for our  Yang-Mills dynamical system:
\begin{eqnarray}
&&\frac{d}{dt}\dot{A}{}^{\ast \alpha i}({\mathbf x},t)+\left(
-2\,c^{\alpha}_{\epsilon \beta}\,\dot{A}{}^{\ast \epsilon  i}({\mathbf x},t)\int d{\mathbf y}\,{\mathscr A}^{(\beta,x)}_{\;\;\;(\sigma,j,y)}\dot{A}{}^{\ast \sigma j}({\mathbf y},t)\right.\nonumber\\
&&+\left.c^{\alpha}_{\mu \beta}\int d{\mathbf y}d{\mathbf z}\,{\mathscr A}^{(\beta,x)}_{\;\;\;(\epsilon,k,z)}
\left[{\cal D}^{\mu i}_{\;\nu}(A^{\ast}({\mathbf x},t))
{\mathscr A}^{(\nu,x)}_{\;\;\;(\sigma,j,y)}
\right]\,
\dot{A}{}^{\ast \sigma j}({\mathbf y},t)\,\dot{A}{}^{\ast \epsilon k}({\mathbf z},t)\right)\nonumber\\
&&+''{\mathscr F} -{\rm terms}''  \nonumber\\
&&\int d{\mathbf u}\,d{\mathbf z}\left(-c^{\alpha}_{\mu \nu}\,{\gamma}^{(\sigma,z)\,(\mu, x)}
\left[{\cal D}^{\nu i}_{\;\epsilon}(A^{\ast}({\mathbf x},t))
{\gamma}^{(\kappa,u)\,(\epsilon, x)}\right]\right.\nonumber\\
&&\;\;\;\;\;\;\left.+c^{\sigma}_{\varphi \mu}\,{\gamma}^{(\mu,z)\,(\kappa, u)}
\left[{\cal D}^{\alpha i}_{\;\epsilon}(A^{\ast}({\mathbf x},t))
{\gamma}^{(\varphi,z)\,(\epsilon, x)}\right]
\right)p_{\kappa}({\mathbf u},t)\,p_{\sigma}({\mathbf z},t)\nonumber\\
&&+G^{(\alpha , i, x)(\gamma,m,y)}\frac{{\delta}}{\delta A^{(\gamma ,m,y)}}\;V[A]\Big|_{A=A^{\ast}}=0,
\label{horyang}
\end{eqnarray}
where  $``{\mathscr F} -{\rm terms}$''  denotes those  terms of the functional representation that correspond to the terms of eq.(\ref{6}) with the curvature ${\mathscr F}^{\nu}_{EF}$.  All these terms  are linear in $p$ and are explicitly given as
\begin{enumerate}
\item
\[
 -2 c^{\sigma}_{\varphi \mu}k^{\beta \alpha}\int d{\mathbf y}d{\mathbf z}\,{\mathscr A}^{(\nu ,z)}_{\;\;\;(\sigma,k,y)}
\,{\mathscr A}^{(\mu,i,y)}_{\;\;\;(\beta,x)}\,\dot{A}{}^{\ast \varphi k}({\mathbf y},t)\,p_{\nu}({\mathbf z},t)
\]
\item 
\begin{eqnarray*}
&&c^{\sigma}_{\mu \epsilon}\int d{\mathbf y}\,d{\mathbf z}\,
\left\{
k_{\sigma\varphi}
\left[{\partial}_{k}({\mathbf y}){\gamma}^{(\nu,z)\,(\epsilon , y)}
\right]
k^{\beta\alpha} \,{\mathscr A}^{(\mu,i,y)}_{\;\;\;(\beta,x)}\right.
\nonumber\\
&&+\left.{\gamma}^{(\nu,z)\,(\epsilon , y)}k_{\rho\varphi}\left[ {\cal D}^{\rho}_{\;\sigma k}(A^{\ast}({\mathbf y},t))\,{\mathscr A}^{(\mu,i,y)}_{\;\;\;(\beta,x)}\right]k^{\beta\alpha}
\right\}\,\dot{A}{}^{\ast \varphi k}({\mathbf y},t)\,p_{\nu}({\mathbf z},t)
\end{eqnarray*}
\item
\[
 2\, c^{\alpha}_{\beta \mu}\left(\int d{\mathbf z}\,{\gamma}^{(\nu,z)\,(\mu , x)}\,p_{\nu}({\mathbf z},t)\right)
\dot{A}{}^{\ast \varphi i}({\mathbf x},t)
\]
\item 
\[
 2 c^{\sigma}_{\rho \mu}k^{\rho \alpha}\int d{\mathbf y}d{\mathbf z}\,{\mathscr A}^{(\nu,i,z)}_{\;\;\;(\sigma,x)}
\,{\mathscr A}^{(\mu,x)}_{\;\;\;(\varphi,k,y)}\,\dot{A}{}^{\ast \varphi k}({\mathbf y},t)\,\,p_{\nu}({\mathbf z},t)
\]
\item 
\begin{eqnarray*}
&&-c^{\sigma}_{\mu \epsilon}\int d{\mathbf y}\,d{\mathbf z}\,
\left\{
{\delta}^{\alpha}_{\sigma}
\left[{\partial}^{i}({\mathbf x}){\gamma}^{(\nu,z)\,(\epsilon , x)}
\right]
 \,{\mathscr A}^{(\mu,x)}_{\;\;\;(\varphi,k,y)}\right.
\nonumber\\
&&+\left.{\gamma}^{(\nu,z)\,(\epsilon , x)}\left[ {\cal D}^{\alpha i}_{\;\sigma }(A^{\ast}({\mathbf x},t))\,{\mathscr A}^{(\mu,x)}_{\;\;\;(\varphi,k,y)}\right]
\right\}\,\dot{A}{}^{\ast \varphi k}({\mathbf y},t)\,p_{\nu}({\mathbf z},t)
\end{eqnarray*}
\item 
\[
 c^{\nu}_{\beta \sigma}k^{\mu \alpha}\int d{\mathbf y}\,d{\mathbf z}\,{\mathscr A}^{(\beta,z)}_{\;\;\;(\varphi,k,y)}
\,{\mathscr A}^{(\sigma,i,z)}_{\;\;\;(\mu,x)}\,\dot{A}{}^{\ast \varphi k}({\mathbf y},t)\,p_{\nu}({\mathbf z},t)
\]
\end{enumerate}
The last term of eq.(\ref{horyang}) is given by
\[
 G^{(\alpha , i, x)(\gamma,m,y)}\frac{{\delta}}{\delta A^{(\gamma ,m,y)}}\;V[A]\Big|_{A=A^{\ast}}=2{\cal D}^{\alpha}_{\;\beta j}(A^{\ast}({\mathbf x},t))(F^{\beta}){}^{i j}({\mathbf x},t).
\]

The vertical equation of motion looks as follows:
\begin{eqnarray}
&&\frac{d}{dt}\,p_{\sigma}({\mathbf x},t)-c^{\kappa}_{\varphi \sigma}\,p_{\kappa}({\mathbf x},t)\,\int d{\mathbf y}\,{\mathscr A}^{(\varphi,x)}_{\;\;\;(\beta,j,y)}\,\dot{A}{}^{\ast \beta j}({\mathbf y},t)\,\nonumber\\
&&-c^{\varphi}_{\sigma \epsilon}\,p_{\varphi}({\mathbf x},t)\int d{\mathbf y}\,{\gamma}^{(\epsilon , x)\,(\mu , y)}
\,p_{\mu}({\mathbf y},t)
=0.
\label{vertyang}
\end{eqnarray}

We see that equations (\ref{horyang}) and (\ref{vertyang}) of the resulting system are rather complicated integro-differential equations. The transition to equilibrium equations is achieved by using the substitution $\dot{A}{}^{\ast \alpha i}({\mathbf x},t)=0$ in these equations.

Thus, to determine the relative equilibrium,  we must solve the following equations:
\begin{eqnarray}
&&\int d{\mathbf u}\,d{\mathbf z}\left(-c^{\alpha}_{\mu \nu}\,{\gamma}^{(\sigma,z)\,(\mu, x)}
\left[{\cal D}^{\nu i}_{\;\epsilon}(A^{\ast}({\mathbf x}))
{\gamma}^{(\kappa,u)\,(\epsilon, x)}\right]\right.\nonumber\\
&&\;\;\;\;\;\;\left.+c^{\sigma}_{\varphi \mu}\,{\gamma}^{(\mu,z)\,(\kappa, u)}
\left[{\cal D}^{\alpha i}_{\;\epsilon}(A^{\ast}({\mathbf x}))
{\gamma}^{(\varphi,z)\,(\epsilon, x)}\right]
\right)p_{\kappa}({\mathbf u})\,p_{\sigma}({\mathbf z})=\nonumber\\
&&-2{\cal D}^{\alpha}_{\;\beta j}(A^{\ast}({\mathbf x}))(F^{\beta}){}^{i j}({\mathbf x})
\label{horequilibr}
\end{eqnarray}
and
\begin{equation}
-c^{\varphi}_{\sigma \epsilon}\,p_{\varphi}({\mathbf x})\int d{\mathbf y}\,{\gamma}^{(\epsilon , x)\,(\mu , y)}
\,p_{\mu}({\mathbf y})
=0.
\label{vertequilibr}
\end{equation}
 The solutions of eq.(\ref{vertequilibr}) are now the eigenfunctions of the Green function ${\gamma}^{(\epsilon , x)\,(\mu , y)}$.
Substitution of these eigenfunctions into eq.(\ref{horequilibr}) leads to an equation for determining the gauge fields characterizing the equilibrium.
\section{Concluding remarks}
The main fundamental difference from the finite-dimensional case is that in a pure Yang-Mills dynamical system there can in principle be a countable number of equilibria. Apparently, this also occurs in other gauge systems. 
But since the quantum vacuum is usually chosen so as to be related to the equilibrium, an important question arises: what role do these   additional equilibria play in the quantization of gauge systems?

For $p_{\mu}=0$, the classical equation (\ref{4}) describes the motion of the reduced system given on the orbit space. In this case, the equation (\ref{equilibr}) (and also (\ref{horequilibr})) for the relative equilibrium  has the standard form of the  equilibrium equation. Is it possible to assume that we are dealing with a description of some  excitation of the system if the ground quantum state is related to one of the relative equilibria obtained for the case $p_{\mu} \neq 0$ ? Is there a connection between the last approach and the quantization of the system obtained as a result of the reduction of the initial system onto the non zero momentum level? 

The answers to these questions arising from the equations of relative equilibrium are very important for our understanding of the classical and quantum behavior of gauge systems.  
Therefore, in subsequent studies, if it is proved that solutions of the equations exist, it would also be desirable to consider the above issues.


\end{document}